\documentclass[aps,prb,twocolumn,showpacs,preprintnumbers]{revtex4-1}
\usepackage{bm,hyperref}
\usepackage{amsmath}
\usepackage{graphicx}
\usepackage{soul}
\usepackage{amsfonts}
\usepackage{amssymb}
\usepackage{dcolumn}
\usepackage{amsmath}

\begin{document}
\title{Optical Force and Torque on Dipolar Dual Chiral Particles}

\author{A. Rahimzadegan{$^{*1}$}, M. Fruhnert{$^{1}$}, R. Alaee{$^{1,2}$}, I. Fernandez-Corbaton{$^{3}$}, and C. Rockstuhl{$^{1,3}$}}
\address{
{$^{1}$}Institute of Theoretical Solid State Physics, Karlsruhe Institute of Technology,D-76131 Karlsruhe, Germany\\
{$^{2}$}Max Planck Institute for the Science of Light, 91058 Erlangen, Germany\\
{$^{3}$}Institute of Nanotechnology, Karlsruhe Institute of Technology, 76344 Karlsruhe, Germany\\
$^*$Corresponding author: aso.rahimzadegan@kit.edu}

\begin{abstract}
On the one
hand, electromagnetic dual particles preserve the helicity of light upon interaction. On the other hand, chiral particles respond differently to light of opposite helicity. These two properties on their own constitute a source of fascination. Their combined action, however, is less explored. Here, we study on analytical grounds the force and torque as well as the optical cross sections of dual chiral particles in the dipolar approximation exerted by a particular wave of well-defined helicity: A circularly 
polarized plane wave. We put emphasis on particles that possess a maximally electromagnetic chiral and hence dual response.  Besides the analytical insights, we also investigate the exerted optical force and torque on a 
real particle at the example of a metallic helix that is designed to approach the maximal electromagnetic chirality condition. Various applications in the context of optical sorting but also nanorobotics can be foreseen considering the particles studied in this contribution.
\end{abstract}
\pacs{ 78.20.Bh, 78.67.Bf, 87.85.St,81.05.Xj}
\maketitle

Electromagnetic (EM) duality is the symmetry of nature for electric and magnetic quantities ~\cite{calkin1965invariance,zwanziger1968quantum,schwarz1999string,mignaco2001electromagnetic} and it can be exploited in the study of light-matter interaction~\cite{FernandezCorbaton2014a}. This symmetry, as implied by the source-free Maxwell's equations, holds in free space, but it is normally broken inside a material. This is because materials respond
differently to electric and magnetic fields, i.e.~the permittivity is different from the permeability. However, dual particles, at least in the dipolar approximation, are realizable~\cite{Kerker:83,zambrana2013dual,fernandez2015dual,Alaee2015a}. Then, it is only required that their electric and magnetic dipole polarizabilities are identical and that the cross polarizabilities are equal in magnitude but opposite in sign ~\cite{fernandez2013role}. This can be achieved even using basic objects such as a sphere made from a material with a sufficiently large permittivity~\cite{Person:2013,Staude:2013}. For a certain combination of diameter, permittivity, and wavelength, the lowest order electric and magnetic Mie coefficient will be the same, satisfying the condition of duality. 

Mathematically, EM duality is the invariance of Maxwell's equations under the non-geometric duality transformation of \cite{Jackson1999}:
  
\begin{eqnarray}
\mathbf{E} & \rightarrow & \mathbf{E} \cos \theta  - Z\mathbf{H} \sin \theta ,\\
Z\mathbf{H} & \rightarrow & \mathbf{E} \sin \theta  + Z\mathbf{H} \cos \theta,
\end{eqnarray}

\noindent where $\theta$ is an arbitrary angle, $Z$ is the impedance of the medium; and $\mathbf{E}$ and $\mathbf{H}$ are the electric and magnetic field vectors, respectively. The importance of dual particles is that they preserve
the helicity of the incident light. Helicity $\Lambda$ is the light's total angular momentum $\mathbf{J}$ in the direction of the light's linear momentum $\mathbf{P}$, i.e.~$\Lambda=\mathbf{J} \cdot \mathbf{P}/|\mathbf{P}|$ \cite{tung1985group}.
 
\begin{figure}
\centering
\includegraphics[clip, trim=1.25cm 0 1.2cm 0, width=0.48\textwidth]{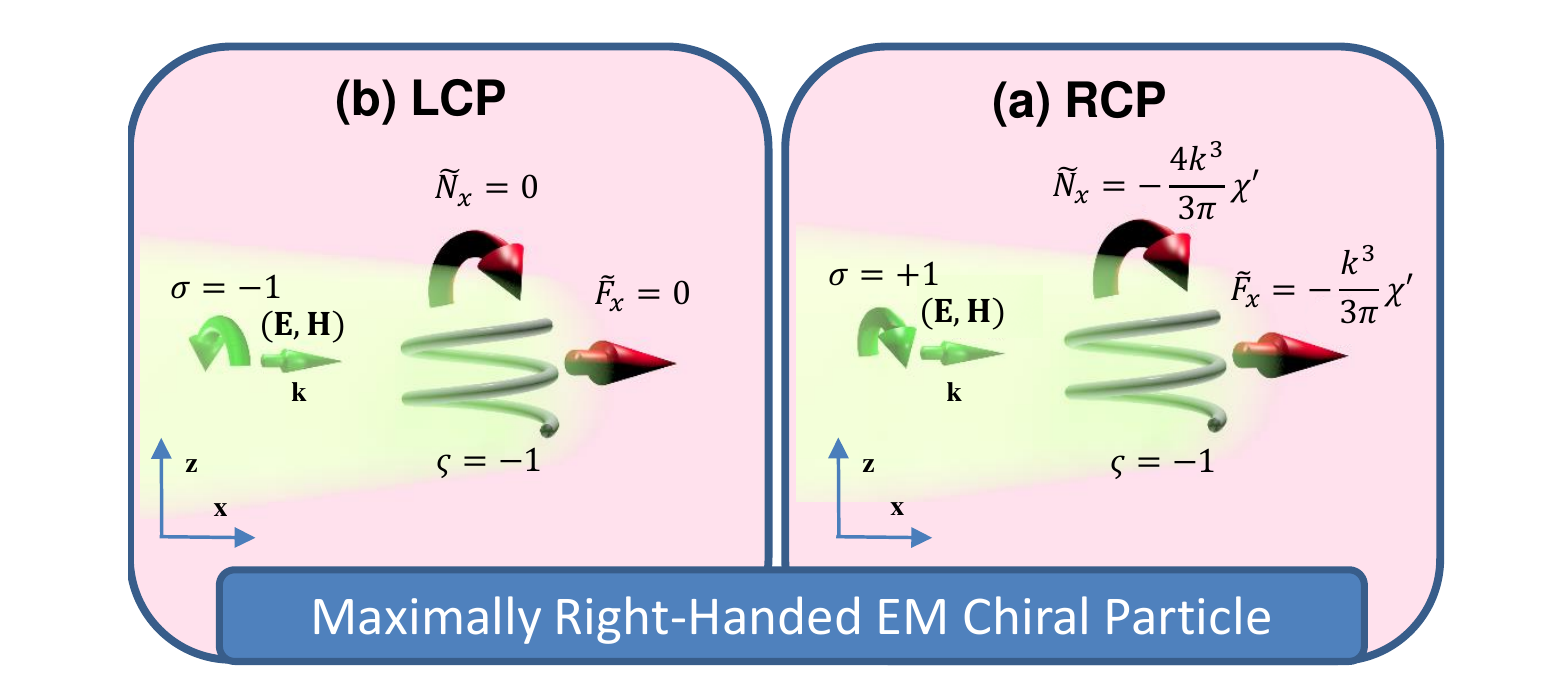}
\caption{The main idea of our work: The normalized time averaged optical force and torque exerted on a maximally right-handed ($\varsigma=-1$) EM chiral particle (a) by a left-handed circularly polarized ($\sigma=-1$) plane wave is zero and b) by a right-handed circularly polarized ($\sigma=+1$) plane wave is maximal.\label{fig:Art}}
\end{figure}

While EM duality necessitates the preservation of the incident light's helicity, chirality allows for a different strength in the response of the particle to light possessing an opposite helicity \cite{hentschel2012three,plum2009metamaterial,Mackay2013,kaschke2016optical,Serd,gansel2010gold,decker2007circular,lobanov2015controlling,von2010three,hannam2014broadband,zhao2010chiral,PhysRevLett.103.103602}. Chirality, as an intrinsic asymmetry in nature, is an important footprint of biological systems \cite{guijarro2008origin,gubitz2006chiral}. An object is said to be chiral, as it was coined by Lord Kelvin, if it is not superimposable onto its mirror image by any rotation or translation \cite{guijarro2008origin}. The necessary asymmetry is the reason for the different response of the particle when illuminated by waves of opposite helicity.

Chirality on its own, however, is only a binary condition. It neither allows ranking particles according to their chirality nor it allows to conclude how chiral a given object is with respect to a maximal chirality it might attain. These issues have been recently solved \cite{PhysRevX.6.031013} by introducing a measure of chirality, defined as the EM chirality. EM chirality has a well defined upper bound. It has been also shown that a maximally EM chiral particle is necessarily dual, and transparent to one of the polarization handedness of the field \cite{PhysRevX.6.031013}. The analysis of such particles will be a scientifically rich exploration. In this contribution, we concentrate on all aspects related to the optical force and torque exerted on such particles.

The sorting of chiral particles that are deposited on a substrate or while being in solution has been the main objective in many analytical studies on the optical force \cite{Hayat2015, wang2014lateral,fernandes2015optical}. In Ref.~\onlinecite{rahimzadegan2016fundamental}, the optical force and torque on dual spheres has been calculated and it was shown that duality leads to a maximal exerted torque. However, the analysis of the optical force and torque exerted on particles that are simultaneously chiral and dual has not yet been reported. Such analysis is timely, since meeting simultaneously the condition of chirality and duality can be realized with tailored particles.  

To close this gap, we study in this contribution analytically the optical cross sections and the exerted optical force and torque on a dual chiral particle and also a maximally EM chiral particle in the dipolar approximation. A metallic helix is designed to approximate the conditions of maximal EM chirality at THz frequencies \cite{PhysRevX.6.031013}. To compare our analytical prediction to numerical simulations, the force and torque exerted by a circularly polarized plane wave on such a helix is simulated using the T-matrix method \cite{fruhnert2016tunable} and the multipole expansion \cite{muhlig2011multipole, Xu1995c, mackowski2008exact}. Figure~\ref{fig:Art} shows the main idea of our work. The optical force and torque on a maximally right-handed EM chiral particle is maximal/zero for a right-/left-handed circularly polarized illumination. The opposite holds of course for particles possessing the opposite handedness. The insights we obtain here are important in the context of optical manipulation.

\paragraph*{Chirality in dipole approximation:}
For a small particle in vacuum, illuminated by a time-harmonic EM wave oscillating at the frequency $\omega$, the induced electric
and magnetic polarizations can be expanded into Cartesian multipole moments. Neglecting higher
orders, the induced electric and magnetic dipole moments ($\mathbf{p}$ and $\mathbf{m}$) can be expressed
as \cite{Serd}:

\begin{eqnarray}
\left[\begin{array}{c}
\mathbf{p}\left(\mathbf{r},\omega\right)/\epsilon_{0}\\
\mathit{Z}_{0}\mathbf{m}\left(\mathbf{r},\omega\right)
\end{array}\right] & = & \bar{\bar{\alpha}}\left(\omega\right)\left[\begin{array}{c}
\mathbf{E}\left(\mathbf{r},\omega\right)\\
Z_{0}\mathbf{H}\left(\mathbf{r},\omega\right)
\end{array}\right],\label{eq:bianisotrop_dipoles-1}\\
\bar{\bar{\alpha}}\left(\omega\right) & = & \left[\begin{array}{cc}
\bar{\bar{\alpha}}^{ee}\left(\omega\right) & \bar{\bar{\alpha}}^{em}\left(\omega\right)\\
\bar{\bar{\alpha}}^{me}\left(\omega\right) & \bar{\bar{\alpha}}^{mm}\left(\omega\right)
\end{array}\right],\label{eq:bianisotropic-polarizability-dipole}
\end{eqnarray}

\noindent where $\mathbf{E}$ and $\mathbf{H}$ are the incident electric
and magnetic fields at the location of the particle $\mathbf{r}$, $\mathit{Z}_{0}$ and $\epsilon_{0}$
are the impedance and permittivity of the free space, and $\bar{\bar{\alpha}}^{ee}$,
$\bar{\bar{\alpha}}^{em}$, $\bar{\bar{\alpha}}^{me}$,
and $\bar{\bar{\alpha}}^{mm}$ are the electric, electric-magnetic,
magnetic-electric, and magnetic polarizability dyadics, respectively.
If the particle's optical response can be approximated by dipole
moments in a certain spectrum, then the particle may be referred
to as a \textit{dipolar particle}. For a static field bias,
the polarizability elements are real. However, for a dynamic EM
excitation, the elements of the polarizability tensors, known as the
\textit{dynamic polarizability tensors}, are complex valued \cite{astapenko2013polarization}. The dipole
polarizability tensor $\bar{\bar{\alpha}}$ has 36 matrix elements
in total. Certain symmetries, implicit in Maxwell's equations, impose
some restrictions on the polarizability tensor. For a reciprocal dipolar
particle, the following Onsager relations hold for both static
and dynamic polarizabilities \cite{Mackay2013,Serd,Sersic2011}:

\begin{equation}
\bar{\bar{\alpha}}^{ee}=\left(\bar{\bar{\alpha}}^{ee}\right)^{\mathrm{T}},\,\bar{\bar{\alpha}}^{mm}=\left(\bar{\bar{\alpha}}^{mm}\right)^{\mathrm{T}},\,\bar{\bar{\alpha}}^{em}=-\left(\bar{\bar{\alpha}}^{me}\right)^{\mathrm{T}},
\end{equation}

\noindent where superscript T denotes the matrix transpose.

If $\bar{\bar{\alpha}}^{em}=-\left(\bar{\bar{\alpha}}^{me}\right)^{\mathrm{T}}\neq\mathbf{0}$,
the particle is bianisotropic. A bianisotropic particle is both
anisotropic and magneto-electric. Magnetoelectrism is the cross-coupling
between the incident electric/magnetic fields and the induced magnetic/electric
multipole moments \cite{Serd,Mackay2013,vehmas2013eliminating}.

The two general categories of bianisotropic materials/particles are
reciprocal and nonreciprocal. Without external bias fields or some
specific cases (i.e moving particles), all materials/particles are reciprocal.  Chiral particles are an important category of bianisotropic particles. We define a dipolar chiral particle in the following way:
\textit{A dipolar particle is electromagnetically chiral if its mirrored dipolar polarizability
tensor $\bar{\bar{\alpha}}\left(\omega\right)$ cannot be brought back to its initial form by any 3D rotation of the tensor.}

In other words, $\bar{\bar{\mathbf{M}}}\bar{\bar{\alpha}}\bar{\bar{\mathbf{M}}}^{-1}$
should not be superposable on itself by any rotation, where $\mathrm{\bar{\bar{\mathbf{M}}}}$
is the mirror operator. A mirror operation is a space inversion and
a $\pi$ rotation with respect to the mirror plane, i.e.~$\bar{\bar{\mathbf{M}}}_{j}=\bar{\bar{\Pi}}\cdot\mathrm{\bar{\bar{R}}}\left(j,\theta=\pi\right),\, j=x,y,z$,
where $\bar{\bar{\mathbf{M}}}_{j}$ is the mirror operator with respect to the $j=0$
plane, $\bar{\bar{\Pi}}$ is the parity operator doing the space inversion $\left(\mathbf{r}\rightarrow-\mathbf{r}\right)$
and $\mathbf{\bar{\bar{R}}}\left(j,\theta=\pi\right)$
is a $\pi$ rotation with respect to the $j=0$ plane. Rotation and
parity operators commute, i.e.~$\bar{\bar{\Pi}}\cdot\mathrm{\mathrm{\mathbf{\bar{\bar{R}}}}}=\mathrm{\mathrm{\mathbf{\bar{\bar{R}}}}}\cdot\bar{\bar{\Pi}}$.
Therefore, in order to prove that a particle is not chiral, it would
be enough to find only one possible rotation $\mathrm{\mathbf{\bar{\bar{R}}}}$
in the three dimensional space to satisfy the expression:

\begin{eqnarray}
\mathbf{\bar{\bar{R}}}\bar{\bar{\alpha}}\left(\omega\right)\mathbf{\bar{\bar{R}}}^{-1}&=&\bar{\bar{\Pi}}\bar{\bar{\alpha}}\left(\omega\right)\bar{\bar{\Pi}}^{-1}.\label{eq:Chirality}\end{eqnarray}

It can be proven that a particle with broken symmetries in three orthogonal
planes is necessarily chiral. However, not all bianisotropic particles are chiral. A particle which is bianisotropic ($\bar{\bar{\alpha}}^{em}\neq0$)
but not chiral is called an omega particle \cite{Serd,Alaee2015reciprocal}. It is called omega because a structural implementation that possesses such a polarizability tensor has the shape of an omega letter ($\Omega$).

\paragraph*{Optical field scattering and duality:}

The scattered light into the far-field of a dipolar particle can be written as \cite{Jackson1999}:

\begin{eqnarray}
\mathbf{E}_{\mathrm{sca}}& = & \frac{Z_{0}k^{2}}{4\pi}\left[\left(\mathbf{n}\times c\mathbf{p}\right)\times\mathbf{n}-\left(\mathbf{n}\times\mathbf{m}\right)\right]\frac{e^{\mathrm{i}kr}}{r},\\
\mathbf{H}_{\mathrm{sca}} & = & \frac{k^{2}}{4\pi}\left[ \left(\mathbf{n}\times\mathbf{m}\right) \times\mathbf{n}+ \left(\mathbf{n}\times c\mathbf{p}\right)\right] \frac{e^{\mathrm{i}kr}}{r},
\end{eqnarray}

\noindent where $\mathbf{n}$ is the unit vector in the direction of the far-field radiation and $k$ is the wavenumber. For simplicity, space and frequency arguments are suppressed from here on. There is an interesting symmetry between the electric and magnetic dipole moments in the equation above. If $c \mathbf{p}=\pm \mathrm{i} \mathbf{m}$, then we have $\mathbf{H}_{\mathrm{sca}} =  \mp \mathrm{i}\mathbf{E}_{\mathrm{sca}}/{Z_{0}}$, which describes a wave with a well-defined helicity. Therefore, for the two conditions, the particle scatters only waves of a well-defined helicity \cite{fernandez2013role,FernandezCorbaton2014a}. In other words, if the particle is illuminated by a wave of well-defined helicity, the scattered field also have a well-defined helicity.  Therefore, satisfying $c \mathbf{p}=\pm \mathrm{i} \mathbf{m}$ is enough to realize EM duality or anti-duality for a dipolar particle for all incident fields \cite{fernandez2015dual,zambrana2013dual}. Here, we will only focus on dipolar dual particles.

\paragraph*{Dual chiral particles:}

Assume a very simplistic representation of a chiral particle that only shows a response to a z-polarized electric or magnetic field. Then, the polarizabilities are as follows:
\begin{eqnarray}
\bar{\bar{\alpha}}^{ee}=\alpha_{zz}^{ee}\mathbf{e}_{z}\mathbf{e}_{z},\,\bar{\bar{\alpha}}^{mm}=\alpha_{zz}^{mm}\mathbf{e}_{z}\mathbf{e}_{z}\nonumber,\\ \bar{\bar{\alpha}}^{em}=-\left(\bar{\bar{\alpha}}^{me}\right)^{\mathrm{T}}=\alpha_{zz}^{em}\mathbf{e}_{z}\mathbf{e}_{z},
\end{eqnarray}
\noindent with $\mathbf{e}_z$ being the unit vector in the z-direction.

For simplicity we will use the following notations:
\begin{equation}
\alpha_{zz}^{ee}=\alpha^{e},\alpha_{zz}^{mm}=\alpha^{m},\alpha_{zz}^{em}=-\alpha_{zz}^{me}=\chi=\chi^{\prime}+\mathrm{i}\chi^{\prime\prime}.\label{eq:ideal_helix}
\end{equation}

Under a circularly polarized plane wave illumination $\mathbf{E}=E_{\mathrm{0}}e^{\mathrm{i}kx}\left(\mathbf{e}_{y}+\mathrm{i}\sigma\mathbf{e}_{z}\right)/\sqrt{2}$,
the dipole moments induced in the particle can be derived as ($\sigma= \pm 1$
is the handedness of the wave):

\begin{eqnarray}
\mathbf{p}=p_{z}\mathbf{e}_{z} & = & \frac{\epsilon_{0}E_{0}}{\sqrt{2}}\left(\chi+\mathrm{i}\sigma\alpha^{e}\right)\mathbf{e}_{z},\label{eq:p}\\
\mathbf{m}=m_{z}\mathbf{e}_{z} & = & \frac{E_{0}}{\sqrt{2}\mathit{Z}_{0}}\left(\alpha^{m}-\mathrm{i}\sigma\chi\right)\mathbf{e}_{z}.\label{eq:m}
\end{eqnarray}

If $\alpha^{m}=\alpha^{e}$, then, depending on the handedness of the wave $\sigma$, we have $\mathbf{p}=\pm \mathrm{i} \mathbf{m}/c$. In other words $\alpha^{m}=\alpha^{e}=\alpha$ is an adequate condition for this reciprocal dipolar particle to be dual. Therefore, we identify the dual chiral particle as follows:

\begin{eqnarray}
p_{z} & = & \frac{\epsilon_{0}E_{0}}{\sqrt{2}}\left(\chi+\mathrm{i}\sigma\alpha\right),\\
m_{z} & = & \frac{E_{0}}{\sqrt{2}\mathit{Z}_{0}}\left(\alpha-\mathrm{i}\sigma\chi\right).
\end{eqnarray}

\paragraph*{Maximally EM chiral particles:}

The response of the chiral particle in Eqs.~\ref{eq:p}-\ref{eq:m} is $\sigma$ dependent and it reacts with different strength to right- and left-handed circularly polarized plane waves.
Let us assume that we want to maximize the recently introcuded EM chirality measure \cite{PhysRevX.6.031013} for the particle, or in other words to make the particle transparent to one helicity and responsive to the other without coupling the two. For the particle in Eq.~\ref{eq:ideal_helix}, if we enforce:

\begin{eqnarray}
\alpha^{e} & = & \alpha^{m}=\varsigma\mathrm{i}\chi,\label{eq:Maximum-chirality-1}
\end{eqnarray}

\noindent with $\varsigma=\pm1$, for one polarization $\sigma$ the dipole moments
totally cancel out, while for the other polarization, the induced
dipole moments are significant. $\varsigma$ can be defined as the right $\left(\varsigma=-1\right)$ or left $\left(\varsigma=+1\right)$ handedness of
the chiral particle. Please note that this handedness is a choice based on the geometrical handedness of the corresponding helix \cite{king2003chirality}. For a maximally right-handed $\left(\varsigma=-1\right)$
EM chiral particle, we have:

\begin{eqnarray}
p_{z} & = & \sqrt{2}\epsilon_{0}E_{0}\chi\delta_{1\sigma},\\
m_{z} & = & \frac{-\sqrt{2}\mathrm{i}E_{0}\chi\delta_{1\sigma}}{\mathit{Z}_{0}}=-\mathrm{i}cp_{z}.\label{eq:duality}
\end{eqnarray}

\noindent where $\delta_{1\sigma}$ is the delta Kronecker {[}$\delta_{1\sigma}=1$
for right-handed circularly polarized plane wave illumination and $\delta_{1\sigma}=0$ for left-handed circularly polarized plane wave illumination{]}. It can be inferred that a \textbf{maximally
right-handed EM chiral particle is only sensitive to a right-handed circularly polarized plane wave
illumination}. In general, for reciprocal objects, to maximize their EM chirality, they should be dual and transparent to one helicity \cite{PhysRevX.6.031013}.

Below, we calculate the optical cross sections and
the exerted force and torque on the dual chiral and maximally EM chiral particle.

\paragraph*{Optical cross sections:}

The scattering, extinction, and absorption cross sections characterize
the fraction of power a particle scatters, extincts, or absorbs from an incident illumination. For
a dipolar particle they are defined as \cite{Novotny2012,alaee2015optical}:

\begin{eqnarray}
C_{\mathrm{sca}} & = & \frac{c^{2}Zk^{4}}{12\pi I_{0}}\left(\left|\mathbf{p}\right|^{2}+\left|\frac{\mathbf{m}}{c}\right|^{2}\right), \\
C_{\mathrm{ext}} & = & -\frac{kc}{2I_{0}}\Im\left(\mathbf{p}^{*}\cdot\mathbf{E}+\mu_{0}\mathbf{m}^{*}\cdot\mathbf{H}\right),\label{eq:Sca_moment-1}\\
C_{\mathrm{abs}} & = & C_{\mathrm{ext}}-C_{\mathrm{sca}}, 
\end{eqnarray}

\noindent where $I_{0}=\left|E_{0}\right|^{2}/(2\mathit{Z}_{0})$ is
the intensity of the incident light and $\mu_0$ is the permeability of the free space. The scattering and extinction cross sections for
the dual chiral particle under the circularly polarized
plane wave illumination ($\mathbf{k}=k \mathbf{e}_x$) are derived as:

\begin{eqnarray}
C_{\mathrm{sca}}	&=&	\frac{k^{4}}{6\pi}\left[\left|\alpha\right|^{2}+\left|\chi\right|^{2}+2\Im\left(\sigma \chi\alpha^{*}\right)\right], \\
C_{\mathrm{ext}}&	=&	k \Im\left(\alpha - \mathrm{i} \sigma \chi \right). 
\end{eqnarray}

The optical cross sections for
the maximally right-handed chiral particle under the circularly polarized
plane wave illumination ($\mathbf{k}=k \mathbf{e}_x$) are derived as:

\begin{eqnarray}
C_{\mathrm{sca}} & = & \frac{2k^{4}}{3\pi}\left|\chi\right|^{2}\delta_{1\sigma}, \\
C_{\mathrm{ext}} & = & -2k\chi^{\prime}\delta_{1\sigma},\label{eq:Sca_moment}\\
C_{\mathrm{abs}} & = & -2k\left[\frac{k^{3}}{3\pi}\left|\chi\right|^{2}+\chi^{\prime}\right]\delta_{1\sigma}. 
\end{eqnarray}

As a conclusion for the maximally right-handed EM chiral particle:
\begin{enumerate}
\item The optical cross
sections are zero for a left-handed polarization and maximal for a right-handed polarization of the illuminating plane wave.
\item The maximal values are related to $\left|\chi\right|^{2}$,$\chi^{\prime}$
and $\frac{k^{3}}{3\pi}\left|\chi\right|^{2}+\chi^{\prime}$ for the
scattering, extinction, and absorption cross sections, respectively.
\item For a maximum absolute value of the handedness-contrast for the absorption
cross section, $-2k(\frac{k^{3}}{3\pi}\left|\chi\right|^{2}+\chi^{\prime})$
should be maximized, which can be realized for an optimum absorption
of the particle.
\end{enumerate}

For a non-absorbing maximally chiral particle, $C_\mathrm{abs}$ is zero. Inspired by previous research, it can be argued that a non-absorptive maximally EM chiral particle can lead to a maximum scattering and extinction cross section by a plane wave incidence of the same handedness \cite{rahimzadegan2016fundamental}.

\begin{figure}
\begin{centering}
\includegraphics[scale=0.6]{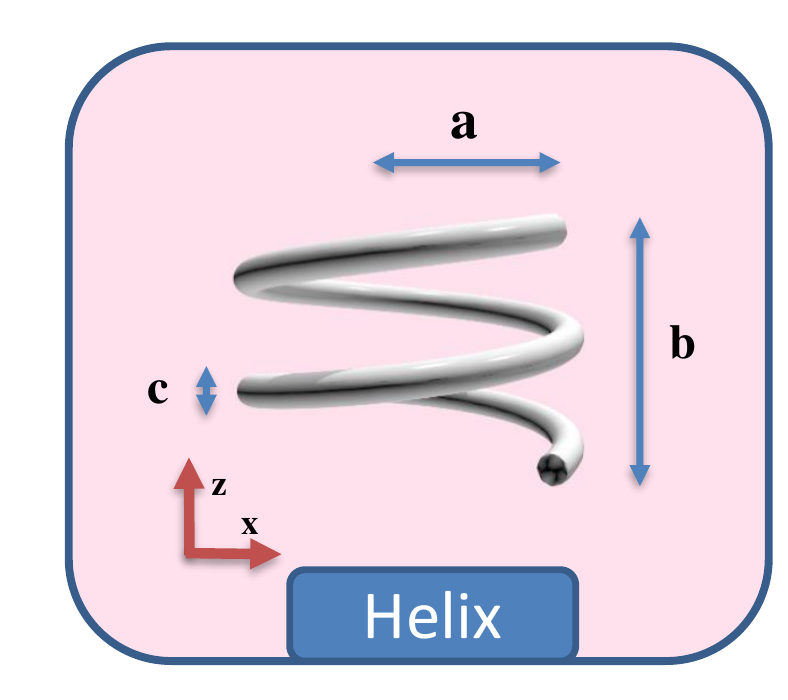}
\par\end{centering}
\protect\caption{The schematic of an optimized maximally right-handed EM chiral helix made of silver with major radius $a=6.48\, \mu$m, height $b=8.52\,\mu$m, and wire radius $c=0.8\, \mu$m.
 {(}Experimental data for silver from Ref.~\onlinecite{hagemann1975optical}{.)} \label{fig:helix}}
\end{figure}

\begin{figure}
\begin{centering}
\includegraphics[scale=0.45]{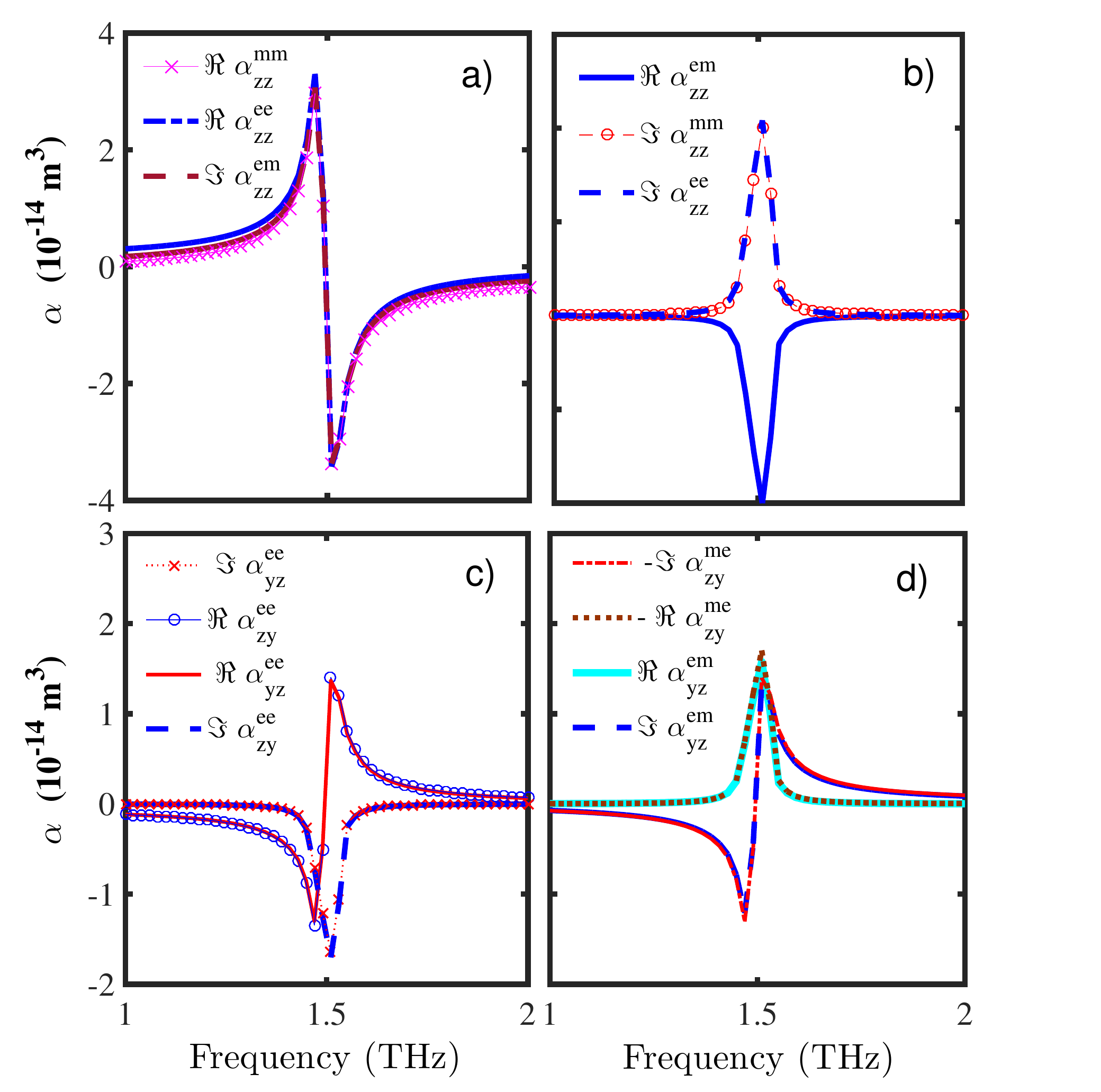}
\par\end{centering}
\protect\caption{The retrieved dynamic polarizabilities of the optimized helix that has been shown in Fig.~\ref{fig:helix}. The maximal EM chirality condition ($\alpha_{zz}^{ee}=\alpha_{zz}^{mm}=-\mathrm{i}\alpha_{zz}^{em}$) is shown in (a) and (b). The reciprocity condition [$\bar{\bar{\alpha}}^{ee}=\left(\bar{\bar{\alpha}}^{ee}\right)^{\mathrm{T}},\,\,\bar{\bar{\alpha}}^{em}=-\left(\bar{\bar{\alpha}}^{me}\right)^{\mathrm{T}}$] is shown in (c) and (d).
\label{fig:polarizability}}
\end{figure}

\paragraph*{Optical force and torque:}

 The optical force and torque can be expressed in terms of multipole moments
of the particle induced by a specific illumination. The multipolar description of force and torque  has been presented in Refs.~\onlinecite{barton1989theoretical,almaas1995radiation}. The results are {\it exact} but lengthy. A full-wave code is developed to calculate the force and torque numerically \cite{rahimzadegan2016fundamental}. The time averaged
optical force and torque exerted on a dipolar particle by an arbitrary incident wave are simplified to (in SI units) \cite{Nieto-Vesperinas2010,National2015a,chen2011optical}:

\begin{eqnarray}
\mathbf{F} & = & \frac{1}{2}\Re\left[\nabla\mathbf{E}^{*}\cdot\mathbf{p}+\nabla\mathbf{B}^{*}\cdot\mathbf{m}-\frac{Z_{0}k^{4}}{6\pi}\left(\mathbf{p}\times\mathbf{m}^{*}\right)\right], \nonumber \\\mathbf{N} & = & \frac{1}{2}\biggl\{\Re\left(\mathbf{p}\times\mathbf{E}^{*}+\mathbf{m}\times\mathbf{B}^{*}\right)\nonumber \\
 &  & -\frac{k^{3}}{6\pi}\left[\frac{1}{\epsilon_{0}}\Im\left(\mathbf{p}^{*}\times\mathbf{p}\right)+\mu_{0}\Im\left(\mathbf{m^{*}}\times\mathbf{m}\right)\right]\biggr\}.\label{eq:Force_moments}
\end{eqnarray}


Under the circularly polarized plane wave $\mathbf{E}=E_{\mathrm{0}}e^{\mathrm{i}kx}\left(\mathbf{e}_{y}+\mathrm{i}\sigma\mathbf{e}_{z}\right)/\sqrt{2}$ illumination, the time
averaged optical force and torque exerted on the dual chiral particle for the
two polarization handedness are derived as:

\begin{eqnarray}
\mathbf{F}& 	=& 	\frac{k^{3}}{6\pi}\left(\mathit{F}_{p}\right)_{\mathrm{max}}\left[\sigma \Im\left(\alpha\right)-\Re\left(\chi\right)\right]\mathbf{e}_{x},\\
\mathbf{N}& 	=& 	\frac{2k^{3}}{3\pi}\left(N_{p}\right)_{\mathrm{max}}\left[\sigma\Im\left(\alpha\right)-\Re\left(\chi\right)\right]\mathbf{e}_{x},
 \end{eqnarray}
\noindent where $\left(F_{p}\right)_{\mathrm{max}}=3 (I_0/c) (\lambda^2/2\pi)$ and $\left(N_{p}\right)_{\mathrm{max}}=3 (I_0/\omega) (\lambda^2/8\pi)$
are the upper bounds for the optical force and torque that a plane wave can exert on an isotropic electric dipolar particle \cite{rahimzadegan2016fundamental}. It can be concluded that for a dual chiral particle, although the helicity is preserved, the exerted torque can only be zero for one of the polarizations when $\sigma \Im\left(\alpha\right)=\Re\left(\chi\right)$, which is the condition for maximal EM chirality. The relations derived for the force and torque are very similar. This symmetry might be a direct consequence of the duality symmetry.

For the {\it ideally} maximally right-handed EM chiral particle the force and torque derive as:
\begin{eqnarray}
\mathbf{F} & = & F_{x}\mathbf{e}_{x}=-\frac{k^{3}}{3\pi}\left(F_{p}\right)_{\mathrm{max}}\chi^{\prime}\delta_{1\sigma}\mathbf{e}_{x},\label{eq:Force}\\
\mathbf{N} & = & \mathcal{N}_{x}\mathbf{e}_{x}=-\frac{4k^{3}}{3\pi}\left(N_{p}\right)_{\mathrm{max}}\chi^{\prime}\delta_{1\sigma}\mathbf{e}_{x}.\label{eq:Torque}
\end{eqnarray}

The optical force and torque for the maximally right-handed EM chiral particle vanishes
totally for $\sigma=-1$, whereas for a right-handed circularly polarized
plane wave illumination $\left(\sigma=+1\right)$, the value depends
on the real part of the chirality factor $\chi^{\prime}$. For simplicity, we define the normalized
optical force and torque as $\widetilde{F}=F/\left(F_{p}\right)_{\mathrm{max}}$
and $\widetilde{N}=N/\left(N_{p}\right)_{\mathrm{max}}$. As a conclusion for the maximally right-handed EM chiral particle:
\begin{enumerate}
\item The optical force and torque are zero for left-handed polarization and maximal for right-handed polarization of the incident plane wave. 
\item The maximal values of the torque and force only depend on $\chi^{\prime}$. Which directly relates it to the optical extinction cross section.
\item Contrary to isotropic particles, where absorption is a necessary condition to exert a torque by a circularly polarized plane wave \cite{rahimzadegan2016fundamental}, an optical torque can be exerted on a non-absorbing dual chiral or maximally EM chiral particle.
\end{enumerate}

 Similar to the case made for the optical cross sections, it can also be argued that for a non-absorbing maximally EM chiral particle, the exerted force and torque can achieve their maximum value.  

\begin{figure}
\begin{centering}
\includegraphics[scale=0.5]{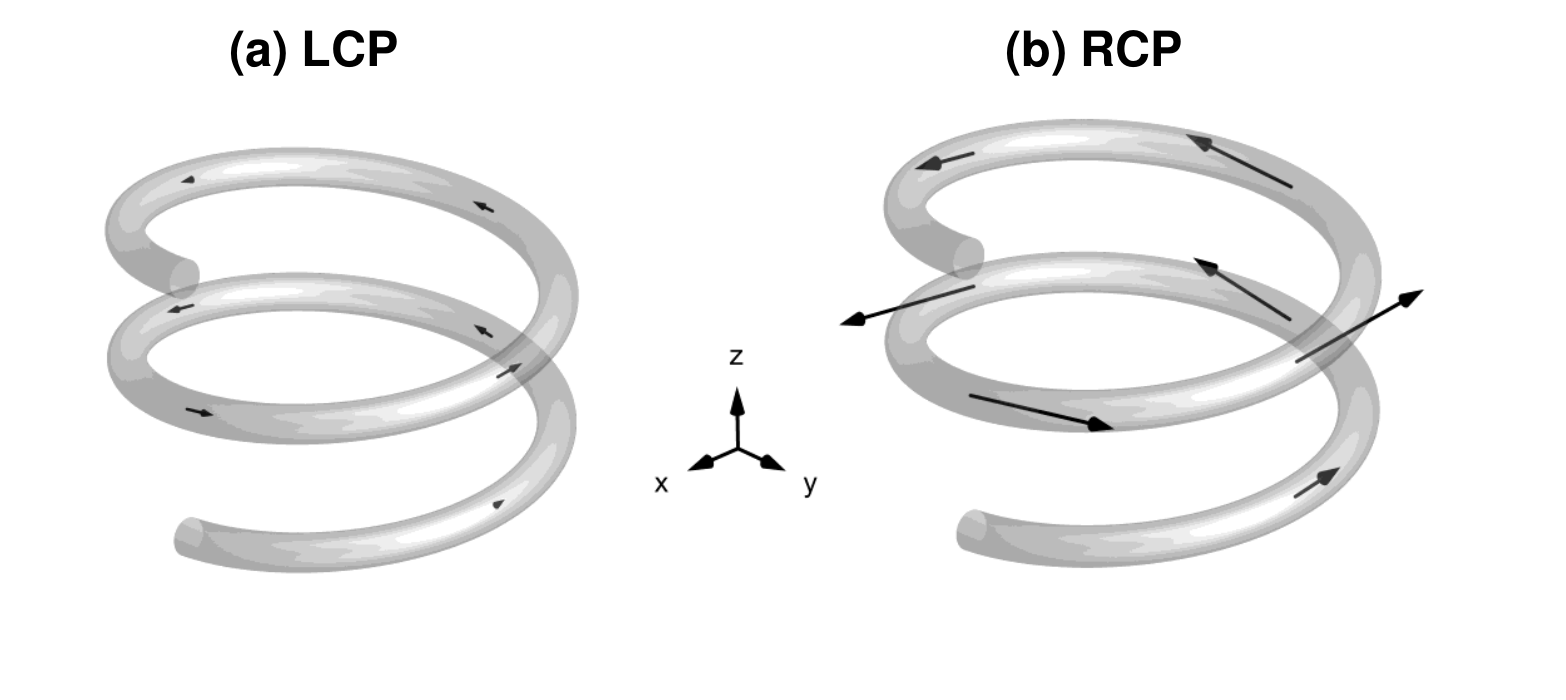}
\par\end{centering}
\protect\caption{The peak current distribution with an arbitrary unit but on the same scale induced in the optimized helix by a (a) left- and (b) right-handed circularly polarized wave illumination. (The calculations are done in COMSOL)
\label{fig:current}}
\end{figure}

\paragraph*{Maximally EM chiral helix:}

Now that we have analyzed the theoretical aspects, we provide
here results of simulations concernig an optimized silver helix in the THz regime. The silver helix has been optimized to satisfy the maximal EM chirality
condition as discussed in Ref.~\onlinecite{PhysRevX.6.031013}. The helix is schematically shown
in Fig.~\ref{fig:helix}. Here, we have retrieved the polarizabilities \cite{alaee2015optical} from the T-matrix of the particle (calculated and discussed in Ref.~\onlinecite{fruhnert2016tunable}) using an existing full-wave code \cite{muhlig2011multipole}. The polarizabilities are in the following form and are shown
in Fig.~\ref{fig:polarizability}:
\begin{eqnarray}
\bar{\bar{\alpha}}^{ee} & = & \alpha_{zz}^{ee}\mathbf{e}_{z}\mathbf{e}_{z}+\alpha_{yz}^{ee}\mathbf{e}_{y}\mathbf{e}_{z}+\alpha_{zy}^{ee}\mathbf{e}_{z}\mathbf{e}_{y},\label{eq:ideal_helix-1}\\
\bar{\bar{\alpha}}^{mm} & = & \alpha_{zz}^{mm}\mathbf{e}_{z}\mathbf{e}_{z},\\
\bar{\bar{\alpha}}^{em} & = & \alpha_{zz}^{em}\mathbf{e}_{z}\mathbf{e}_{z}+\alpha_{yz}^{em}\mathbf{e}_{y}\mathbf{e}_{z},\\
\bar{\bar{\alpha}}^{me} & = & \alpha_{zz}^{me}\mathbf{e}_{z}\mathbf{e}_{z}+\alpha_{zy}^{me}\mathbf{e}_{z}\mathbf{e}_{y}.
\end{eqnarray}

There are non-desirable nonzero elements as shown in the figure, i.e.~those related to a response for a y-polarized field. This might cause some deviations between the excact numerical
results and the analytical predictions in Eqs.~\ref{eq:Force}-\ref{eq:Torque}. Nevertheless, it can be seen that the reciprocity conditions hold
perfectly and the maximal dipolar EM chirality condition (Eq.~\ref{eq:Maximum-chirality-1}) is also approximately 
met. The current distribution on the surface of the helix is also calculated for the two illuminations and is shown in Fig.~\ref{fig:current}. The excitation frequency correspond to the resonance, i.e.~1.5 THz.
The induced current is very small for the left-handed circularly polarized illumination, demonstrating the different response to light with different circular polarization.

\begin{figure} 
\begin{centering}
\includegraphics[width=8cm]{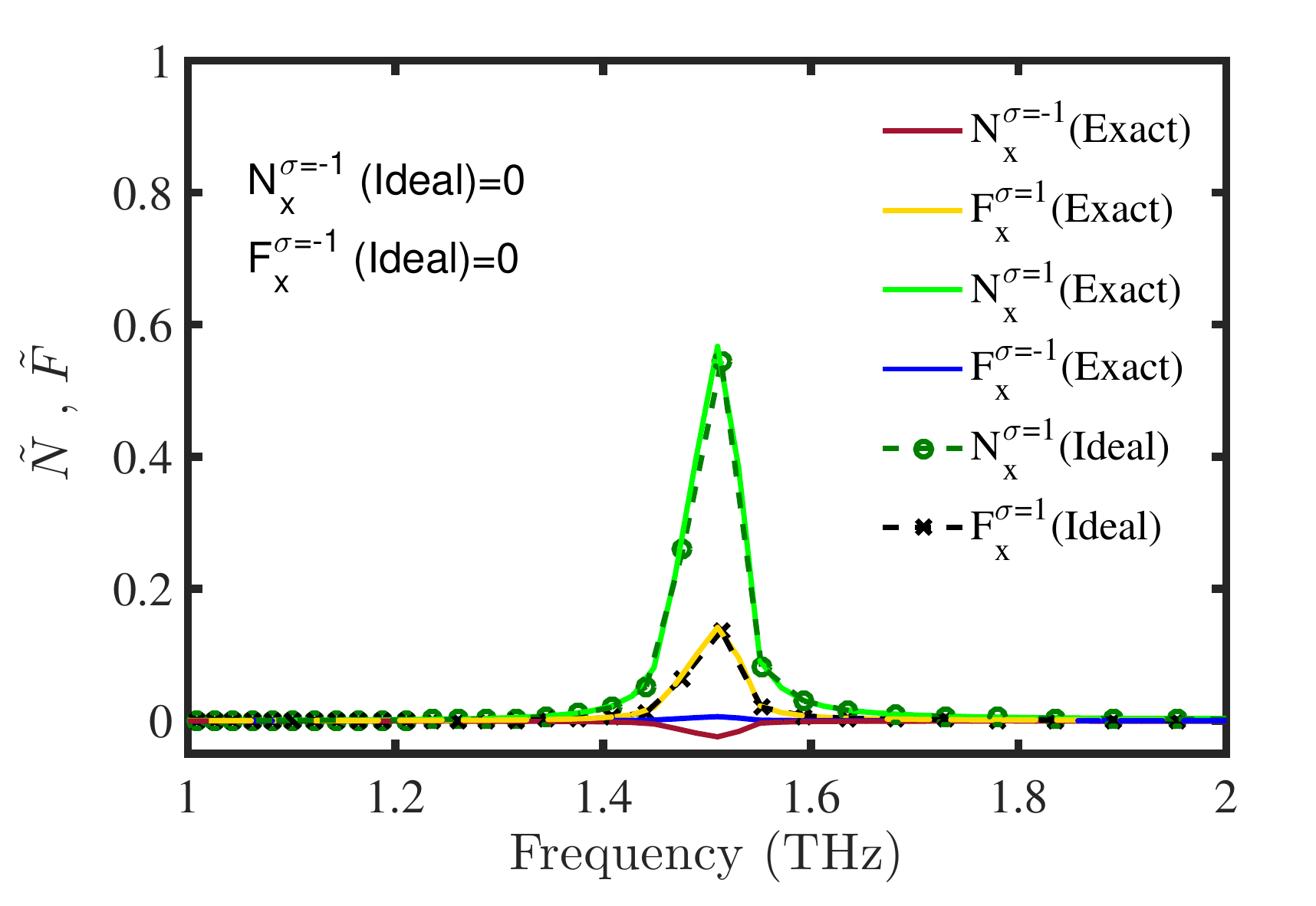}
\par\end{centering}
\protect\caption{The normalized time averaged optical force and torque exerted
on the optimized helix by a circularly polarized plane wave of two
different helicities ($\sigma=\pm1$) propagating in a direction perpendicular
to the helix axis. \label{fig:Force-helix-cp}}
\end{figure}

The {\it exact} force and torque exerted on the helix by a circularly polarized plane wave were calculated numerically using the retrieved T-matrix and the multipole expansion of the optical force and torque \cite{barton1989theoretical,almaas1995radiation}. The numerical calculations are done using a full-wave code ~\cite{rahimzadegan2016fundamental}. Results are shown in Fig.~\ref{fig:Force-helix-cp}. The analytical results asuming that the helix was an {\it ideal} dipolar maximally EM chiral object are plotted using the retrieved cross term polarizability ($ \alpha_{zz}^{em}=\chi$) and the analytical Eqs.~\ref{eq:Force}-\ref{eq:Torque}. The agreement between the exact and ideal results is almost perfect. A non-zero but rather small force and torque in the exact numerical simulation can be seen upon illuminating the right-handed particle with a left-handed circularly polarized plane wave. It is attributed to the fact that the helix is not a perfect maximally EM chiral dipolar object, as can be for example appreciated in the finiteness of the already mentioned entries in the polarizability matrix.

\paragraph*{Conclusion:}
In summary, we have calculated the optical force and torque on dual chiral and maximally EM chiral particles. We have simulated the optical force and torque on a designed maximally EM chiral silver helix. It is shown that a maximally right-handed EM chiral particle can be maximally rotated or accelerated by a right-handed circularly polarized plane wave and become almost transparent to a left-handed circularly polarized plane wave. Of course, the opposite holds for a maximally left-handed EM chiral particle. Our study will be useful in two perspectives. On the one hand, it can contribute to the current research on the optical sorting of racemic mixtures. On the other hand, it can prove useful in optically driven nanorobots. A maximally EM chiral particle can be used as a building block of a robot which is only sensitive to one polarization.

\paragraph*{Acknowledgements:}
We acknowledge partial financial support by the Deutsche Forschungsgemeinschaft through CRC 1173. A.R. and M.F. also acknowledge support from the Karlsruhe School of Optics and Photonics (KSOP).

%


\end{document}